\documentclass[sigconf,screen]{acmart}
\usepackage{soul}

\newcommand{\materialhl}[1]{\colorbox{blue!10}{#1}}
\newcommand{\numpar}{\textsc{15}}

\definecolor{hlBlue}{HTML}{CDE8FF}
\definecolor{hlPink}{HTML}{FFCDD2}
\definecolor{hlGreen}{HTML}{C8E6C9}

\newcommand{\hlb}[1]{{\sethlcolor{hlBlue}\hl{#1}}}
\newcommand{\hlp}[1]{{\sethlcolor{hlPink}\hl{#1}}}
\newcommand{\hlg}[1]{{\sethlcolor{hlGreen}\hl{#1}}}

\usepackage{tabularx}
\newcolumntype{L}{>{\raggedright\arraybackslash}X}

\usepackage[normalem]{ulem}
\useunder{\uline}{\ul}{}
\usepackage{float}

\AtBeginDocument{%
  }

\copyrightyear{2025}
\acmYear{2025}
\setcopyright{cc}
\setcctype{by}
\acmConference[ASSETS '25]{The 27th International ACM SIGACCESS Conference on Computers and Accessibility}{October 26--29, 2025}{Denver, CO, USA}
\acmBooktitle{The 27th International ACM SIGACCESS Conference on Computers and Accessibility (ASSETS '25), October 26--29, 2025, Denver, CO, USA}
\acmDOI{10.1145/3663547.3746393}
\acmISBN{979-8-4007-0676-9/2025/10}

\usepackage{algorithm}
\usepackage{algorithmicx}
\usepackage{algpseudocode}
\usepackage{booktabs}
\usepackage{graphicx}
\usepackage{colortbl}

\begin{document}

\title{Surfacing Variations to Calibrate Perceived Reliability of MLLM-generated Image Descriptions}

% \title{Revealing Variations to Calibrate Perceived Reliability of MLLM-generated Image Descriptions}


\author{Meng Chen}
\affiliation{%
  \institution{The University of Texas at Austin}
  \city{Austin}
  \state{Texas}
  \country{USA}}
\email{mengchen@utexas.edu}

\author{Akhil Iyer}
\affiliation{%
  \institution{The University of Texas at Austin}
  \city{Austin}
  \state{Texas}
  \country{USA}}
\email{akhil.iyer@utexas.edu}

\author{Amy Pavel}
\affiliation{%
  \institution{University of California, Berkeley}
  \city{Berkeley}
  \state{California}
  \country{USA}}
\email{amypavel@eecs.berkeley.edu}

% \begin{teaserfigure}\includegraphics[trim=0cm 0cm 0cm 0cm, clip=true, width=\textwidth]{figures/teaser.pdf}
%     \caption{Variations.}
%     \label{fig:teaser}

\begin{abstract}
Multimodal large language models (MLLMs) provide new opportunities for blind and low vision (BLV) people to access visual information in their daily lives. However, these models often produce errors that are difficult to detect without sight, posing safety and social risks in scenarios from medication identification to outfit selection. While BLV MLLM users use creative workarounds such as cross-checking between tools and consulting sighted individuals, these approaches are often time-consuming and impractical. We explore how systematically surfacing variations across multiple MLLM responses can support BLV users to detect unreliable information without visually inspecting the image. We contribute a design space for eliciting and presenting variations in MLLM descriptions, a prototype system implementing three variation presentation styles, and findings from a user study with 15 BLV participants. Our results demonstrate that presenting variations significantly increases users' ability to identify unreliable claims (by 4.9x using our approach compared to single descriptions) and significantly decreases perceived reliability of MLLM responses. 14 of 15 participants preferred seeing variations of MLLM responses over a single description, and all expressed interest in using our system for tasks from understanding a tornado's path to posting an image on social media.
\end{abstract}





\maketitle

\section{Introduction}
Recent progress in generative AI including multimodal large language models (MLLMs) provides a transformative opportunity for millions of blind and low vision (BLV) people to access both digital and real-world visual information~\cite{WHO_VisualImpairment_2012, adnin2024look, gonzalez2024investigating}.
MLLMs are deep learning algorithms that can process and generate multiple types of content, including text, images, audio, and video ~\cite{Yin_2024}.
BLV people can now use MLLMs through services such as Be My AI ~\cite{BeMyAI}, Seeing AI~\cite{SeeingAI}, and Envision~\cite{EnvisionAI} to support daily activities such as exploring new spaces ~\cite{chang_worldscribe_2024,kuribayashi_wanderguide_2025},
reading informative charts and diagrams, interpreting images on social media posts ~\cite{bennett2018teens},
or creating visual art~\cite{huh2023genassist}. 
MLLMs provide descriptions faster than human-powered access tools~\cite{aira, vizwiz,bemyeyes} and with more detail than traditional image captioning techniques~\cite{huh2024long}. 

While MLLMs produce fluent and persuasive responses, their responses can be factually incorrect or misleading~\cite{huh2024long}. 
For example, MLLMs erroneously \textit{fabricate} content that is not in the image (e.g., state an empty frame contains a family picture), \textit{misinterpret} content that is in the image (e.g., mistaking a cleaning product for shampoo, or \textit{``6mg''} for \textit{``8mg''}), or \textit{omit} important content (e.g., omitting a warning label from a medication description). 
MLLMs also provide overly certain responses for ambiguous queries. For example, a model may confidently report that an outfit matches and is business-casual when humans would disagree.
However, it can be challenging for BLV MLLM users to detect errors or subtle model biases without visually comparing the image to the model response.

BLV MLLM users have thus developed creative strategies to check AI-generated image descriptions such as cross-checking descriptions across multiple tools that use different models, using other senses to verify the response in real-world environments, and asking sighted people for verification~\cite{alharbi2024misfitting}. However, checking across tools or coordinating with sighted people can be time-consuming, and non-visual senses can primarily support users in real-world scenarios. 
Prior work has supported BLV users to check the quality of prior AI-generated descriptions with interactive spatial descriptions~\cite{Nair2023ImageAssist,Lee2022ImageExplorer} or multiple answers to similar questions~\cite{huh2023genassist} that can incidentally reveal model inconsistencies.
% \amy{not really saying why this is bad}
For single descriptions, prior work used the probability of a model response to provide its confidence to the user via natural language or numerical framing (e.g., \textit{``there's a small chance I could be wrong...''}, \textit{``there's a 20\% chance I could be wrong...''})~\cite{macleod2017understanding}.
However, this approach used explicit model confidence scores that are unavailable in ``black-box'' MLLMs, it cannot address the mixture of both reliable (e.g., \textit{``a bar chart''}) and unreliable information (e.g., \textit{``the highest price is \$284,359''}, \textit{``the chart looks polished''}) contained in paragraph-long MLLM descriptions~\cite{huh2024long}.
While BLV users want to be able to assess MLLM responses, and have developed practices of comparing multiple AI-generated responses to do so, no prior work supports BLV users in efficiently generating and comparing MLLM responses. 

Our work aims to support BLV users to detect unreliable information within long-form MLLM responses and foster appropriate perceived reliability of MLLM answers. Our core approach, inspired by BLV users' existing practice of checking multiple tools~\cite{alharbi2024misfitting} and prior literature suggesting that sharing variations can calibrate trust for LLM responses~\cite{lee_one_2024}, is to make it easier for BLV users to generate and compare variations across model responses to surface unreliable visual information. We outline a design space for eliciting and displaying variations in responses produced by black-box MLLMs to support users in understanding the range of possible model responses. From this design space, we prototype three presentation styles: a list of multiple descriptions, a variation-aware description that integrates variations, and a variation summary that explicitly highlights agreements, disagreements, and unique mentions.

To evaluate the effectiveness of surfacing variations in LLM responses, we conducted a controlled user study with 15 BLV participants who regularly used MLLM descriptions comparing a single description, with a list of multiple descriptions, and our approach (a combination of the variation summary and variation-aware description). Our study demonstrates that presenting variations significantly increases users' ability to identify unreliable claims (by 4.9x using our approach compared to a single description) and significantly decreases perceived reliability of MLLM-generated image descriptions. 
Participants preferred our aggregated variation approaches over traditional multiple description lists or single descriptions, with 11 of 15 ranking our variation summary in particular as their favorite option (over the variation-aware descriptions, list of variations, and single descriptions).
All participants expressed interest in using our variation surfacing prototype for future tasks ranging from high-stakes scenarios (e.g., tracking an incoming tornado) to obtaining subjective critiques (e.g., to post an image on social media).

In summary, we contribute: 
\begin{itemize}
    \item A design space for surfacing variations in MLLM-generated image descriptions informed by prior literature
    \item A prototype system that automatically generates and presents variations in MLLM responses in three different presentation styles tailored to BLV users' needs
    \item Empirical findings demonstrating that surfacing variations significantly improves BLV users' ability to identify unreliable information in MLLM responses and decreases their perceived reliability of MLLM responses
\end{itemize}

\section{Related Work}

\subsection{Visual Access Technology}

BLV people use visual assistive technologies to understand visual content in both the real and digital worlds~\cite{adnin2024king}. 
Traditional visual assistance (e.g., Be My Eyes ~\cite{bemyeyes}, VizWiz ~\cite{vizwiz}, Aira ~\cite{aira}) employ sighted human assistants to describe visual content that the user is showing on their camera, but human assistance is not always available~\cite{avila2016remote}. As a result, AI-powered, particularly MLLM-powered (e.g. Be My AI, Meta RayBan), access technologies has become scalable, on-demand alternatives to traditional ones. BLV people use such tools to identify specific objects, build an understanding of scenery, read text and numbers, and identify object locations ~\cite{gonzalez2024investigating, chen_visimark_2025, chang_worldscribe_2024, gubbi2024context}. 
Recent MLLMs can generate long-form answers that include explanations, context, and additional details in response to visual queries ~\cite{huh2024long}. While this nuanced information can help users better understand visual content, the length of these descriptions poses extra challenges. 
First, much like human-written descriptions that can vary in focus and subjective opinions ~\cite{vizwiz, huh2024long}, MLLMs also present information from a specific perspective based on what they choose to describe and how they present it. Second, these detailed outputs often contain small, objective errors known as hallucinations that can be difficult to detect in long-form descriptions, especially when they are otherwise correct (e.g., getting a single number on a chart wrong).
Prior work on traditional captioning models shows that BLV people tend to overly trust AI-generated image descriptions on social media ~\cite{macleod2017understanding,morris_ai_2020} and can only identify about half of the errors when using object detection ~\cite{hong2024understanding}.
MLLM-generated image descriptions are particularly prone to give fluent but incorrect answers, and BLV users are more inclined to perceive such MLLM responses as plausible compared than sighted evaluators ~\cite{huh2024long}. 
In this work, we designed interventions aimed at calibrating BLV users' trust in long-form MLLM-generated image descriptions.

\subsection{Visual Description Verification Strategies for BLV People}
BLV people are early adopters of AI technologies ~\cite{bigham_learning_nodate}, but assessing visual descriptions' correctness remains challenging, as they cannot directly visually verify the descriptions against the image. That said, BLV people have developed multiple verification workarounds in different scenarios. As BLV people use AI-powered visual access tools in everyday tasks, they build up an understanding of errors that tools are prone to make and then check for common object detection errors ~\cite{huh2023avscript, alharbi2024misfitting, gonzalez2024investigating}. Incongruent contextual cues can also suggest potential errors in the description ~\cite{abdolrahmani2017embracing, gonzalez2024investigating}. For example, when \textit{``accordion''} appears in a kitchen scene, BLV people recognize it as an error  ~\cite{ning2024spica}. BLV people often cross-check inconsistency in image descriptions by employing other senses, retaking photos from different angles or with altered backgrounds, or running the same image through multiple apps to compare results and identify inconsistencies~\cite{alharbi2024misfitting, huh2023genassist, hong2024understanding}. BLV people often choose to check with sighted people in high-stakes tasks or scenarios that require accuracy and security ~\cite{zhao2018face, alharbi2024misfitting}.

Prior work supports BLV users in verifying and contesting AI-generated outcomes by eliciting information via question answering ~\cite{chang_editscribe_2024, huh2023genassist} or multi-layered descriptions to enable assessing visual content by checking congruency and consistency ~\cite{ning2024spica, Lee2022ImageExplorer,  Nair2023ImageAssist}.
GenAssist and EditScribe both use variations in generated image descriptions to support BLV creators assessing generated images, such work does not address that the image descriptions themselves may contain errors though repeated descriptions inadvertently surfaced consistency errors~\cite{huh2023genassist}. 
Multi-layered and spatial descriptions~\cite{ning2024spica, Lee2022ImageExplorer,  Nair2023ImageAssist} provide opportunities to explore images and videos spatially and/or hierarchically and thus support BLV users to uncover consistency and incongruence issues in descriptions. 
Building upon the existing visual description verification practices of BLV people, our work reveals variations in multiple MLLM-generated image descriptions from different sources. 
The inconsistencies in MLLM responses can serve as indicators of potential errors that raise the skepticism of BLV users in MLLM-generated image descriptions.

\subsection{Variations in Large Language Model Outputs}
Prior work has explored surfacing variations in large language model (LLM) responses primarily for sighted people. 
Similar to MLLMs, LLMs can generate various outputs given the same input by sampling from a probability distribution of learned words and phrases. Their stochastic characteristics sometimes lead to inconsistent outputs that negatively impacts the safe use of LLMs ~\cite{amodei2016concrete}. While prior machine learning research attempts to develop methods to evaluate \cite{tanneru2024quantifying, ribeiro2019red, asai2020logic, elazar2021measuring, zhang_effect_2020} and reduce these inconsistencies \cite{kellner_uncertainty_2024, wang2022self, cohen_lm_2023}, variations across outputs cannot be eliminated due to the probabilistic nature of LLMs. Therefore, effectively communicating variations and uncertainty to users is critical for safe use. Presenting only a single polished and confident output can mislead users to over-rely on the system and ignore potential errors so prior work work shows that providing explanations~\cite{vasconcelos_explanations_2023}, token probabilities~\cite{vasconcelos_generation_2024}, and multiple variations~\cite{lee_one_2024} can expose AI limitations and mitigate overreliance. 
While prior work has revealed token probabilities directly for code generation~\cite{vasconcelos_generation_2024}, Kuhn et al. \cite{kuhn_semantic_2023} demonstrate that clustering semantically similar responses provides a more meaningful measurement of uncertainty as oppose to traditional metrics like log probability, which often fail to capture semantic nuances \cite{kadavath_language_2022} (e.g., articles like \textit{``the''} and \textit{``a''} may recieve high probability but they are not semantically meaningful).

HCI researchers have thus developed interfaces to visualize these output variations to help users better understand and assess the generated content. Researchers have introduced interactive diagrams \cite{graphologue, 10.1145/3586183.3606756, arawjo2024chainforge}, text renderings \cite{gu_ai-resilient_2024, cheng2024relic}, and multi-output visualizations \cite{gero_supporting_2024, luminate} to make variations more evident. 
However, these interfaces are designed for sighted users first and thus represent variations via visualization techniques like saliency, color codes, placement, and interactive graphics, which are inaccessible to BLV users. For example, Luminate ~\cite{luminate} displays generated variations in an interactive graph visualization.
In this work, we draw on such prior work to design, develop, and evaluate approaches for screen reader users to surface such variations in MLLM rather than LLM responses.
\begin{table*}[!ht]
\centering
\resizebox{5.5in}{!}{%
  \begin{tabular}{@{}lllll@{}}
  \toprule
  \textbf{Dimension}               & \textbf{Alternatives}       &  &  &  \\ 
  \midrule
  \textbf{Elicitation of Variants} & \cellcolor{blue!10}Trials   & \cellcolor{blue!10}Prompts & \cellcolor{blue!10}Models   & Images \\
  \textbf{Comparison Support}      & \cellcolor{blue!10}List & \cellcolor{blue!10}Variation-aware description & \cellcolor{blue!10}Variation Summary & \\
  \textbf{Comparison Granularity}  & Words                       & \cellcolor{blue!10}Atomic facts & Sentences & \cellcolor{blue!10}Responses \\
  \textbf{Support Indicator}     & \cellcolor{blue!10}None     & \cellcolor{blue!10}Percentage & \cellcolor{blue!10}Language & \cellcolor{blue!10}Source \\
  \textbf{Provenance Indicator}    & None                        & Trials         & Prompts        & \cellcolor{blue!10}Models \\
  \textbf{Modality}                & \cellcolor{blue!10}Text     & Sound          & Visualization  & Haptics \\
  \bottomrule
  \end{tabular}%
}
\caption{Design space of surfacing MLLM variations. Colored squares indicate options featured by our prototype. }
\label{tab:design-space}
\end{table*}

\section{Prototype Design and Development}
We aimed to build a system to support BLV users in identifying unreliable information in MLLM-generated responses and assessing model reliability. Towards this goal, we share three design goals based on prior work, create a design space of how systems may support these goals, and share our prototype that instantiates several features of our design space: \\

\noindent \textbf{DG1: Surface variations in MLLM responses. } BLV users currently check model responses across multiple AI tools to assess response accuracy~\cite{alharbi2024misfitting}. Informed by their current practice, and prior work that indicates exposing sighted users to multiple LLM responses reduces trust~\cite{lee_one_2024}, we seek to support BLV users easily surfacing and comparing multiple responses from MLLMs at once. \\ 
% \amy{not well written but ideas are there} \\ 

\noindent \textbf{DG2: Support efficient comprehension of variations. } One approach to surface variations is to simply run an MLLM multiple times then list out all of the variations (e.g., as prior work explored for LLMs~\cite{cheng2024relic}). However, modern MLLM image descriptions and visual question answers are sentences to paragraphs long~\cite{huh2024long} such that it can be cognitively demanding to read and compare variations of the same responses, especially for screen reader users who read the responses linearly. Thus, we seek to create an interface that supports users efficiently understanding variations in model responses. \\

\noindent \textbf{DG3: Support personalized display of variations. } The experience of disability is highly individual and thus accessible technologies must be designed with individual differences in mind~\cite{herskovitz2023hacking}. We seek to support users to customize our tool to meet their needs.  \\ 

\subsection{Surfacing MLLM Variations: A Design Space}
We outline opportunities for systems to support efficiently surfacing MLLM variations in a design space informed by prior work (Table~\ref{tab:design-space}) and explain the dimensions of potential support below. For each dimension, we share what options we included in our prototype for surfacing MLLM variations.

\subsubsection{Elicitation of Variants} Variation in MLLM responses to image and prompt pairs arises from a variety of sources. First, MLLMs are non-deterministic such that there is built in randomness in responses with the same model and a fixed input. Thus, one way to elicit variation is simply to run the same model multiple times with the same input (i.e., to run multiple \textbf{trials}). Second, model interpretations of images are sensitive to prompt variations such as instructing the model to not hallucinate~\cite{huh2024long}, providing a persona~\cite{benharrak2024writer}, or simply paraphrasing the prompt~\cite{ribeiro2019red}. For example, Gemini's response shifted from saying \textit{``the clothes don't really match''} to describing the outfit as \textit{``a fairly standard casual combination''} when the question changed from \textit{``Do they match?''} to \textit{``Do they go well with each other?''} Thus, we can elicit variations by varying the \textbf{prompts} between trials. Finally, different models have different strengths, weaknesses, and patterns of responses (e.g., using hedging or straightforward language~\cite{huh2024long}) thus we can elicit variations by running multiple ~\textbf{models}. Our prototype provides all three options for elicitation and allows users to customize their elicitation strategy. As a default, we use 3 models with 3 trials all with the same prompt to identify meaningful variations without introducing potential confusion from prompt changes or overwhelm from too many responses.  

\subsubsection{Comparison Support} To support people comparing variations, the simplest approach is to \textbf{list} various outputs side by side~\cite{gero_supporting_2024, ChatGPT, lee_one_2024} which provides users full knowledge of the variations but has high cognitive demand to remember the variations among responses. Another approach is to align variations of MLLM responses to a single response that we call a \textbf{variation-aware description} (e.g., ``the chair is red'' becomes ``the chair is \textit{red}, \textit{pink}, or \textit{magenta}''). Prior work explored visualizing such aligned responses for sighted users evaluating LLM responses~\cite{cheng2024relic}. Aligning responses makes it possible for users to understand the content of the image while alerting them to the parts of the description that may be unreliable. However, this approach lengthens the primary description and thus may make it more difficult to understand the image as a whole. A final approach is to create a \textbf{variation summary} that highlights the similarities (e.g., \textit{``all models describe a cat riding a bike''}) and differences (e.g., \textit{``GPT-4V describes the cat as longhair whereas Gemini says the cat is short-hair''}) between MLLM responses. Prior work explored a similar approach to alert BLV users to variations in generated images~\cite{huh2023genassist}. This approach highlights variations most directly, but loses similarity to the original response. We provide all three options in our prototype.

\subsubsection{Comparison Granularity} To compare responses, different levels of granularity may be valuable depending on the task. For example, if the user is trying to compare the output of two MLLM responses reading text to memorize a written poem, they may want \textbf{word-level} comparison such that they can detect any word that varies between the two responses. On the other end of the spectrum, if a BLV user wants to create alt text for their travel photo they may want to read the full responses and select the best one \textbf{response-level}. We also explore two additional points on the spectrum: \textbf{sentence-level} comparison as it provides a more manageable length but often contain multiple pieces of information such that they can be difficult to align between responses, and \textbf{atomic facts} or self-contained units of information (e.g. \textit{``a short-hair cat''}). In our prototype, we use full responses for the list of multiple descriptions and atomic facts for the variation-aware description and variation summary to surface content differences rather than lexical differences in responses. 

\subsubsection{Support indicator} When we present variations to users (e.g., ``the chair is \textit{red}, \textit{pink}, or \textit{magenta}.''), we may surface only unique variations without indicating level of agreement (\textbf{none}) or users may want to know to what extent other models agree with the statement: ``the chair is \textit{red} (90\% of responses), \textit{pink} (6\% of responses), or \textit{magenta} (4\% of responses).'' Similar to how prior work presented model confidence~\cite{macleod2017understanding}, we can present agreement with \textbf{percentage}, natural \textbf{language} indicators (e.g., ``well-supported''), or simple counts of \textbf{source}. We provide all options in our prototype. As a default, we use counts. 

\begin{figure}[htb!]
    \centering
    \includegraphics[alt={A flowchart illustrates a multi-model language processing pipeline for generating image descriptions. The process starts with two inputs: an "Image" and a "Prompt." Both feed into a block labeled "m MLLMs" (e.g., GPT-4o, Gemini 1.5 pro, Claude 3.7 Sonnet), with a parameter "n Sample Size." An optional path also includes "Paraphrased Prompts" as input. The MLLMs output "m×n Descriptions," which are used in two paths. One path leads to a "List View of Descriptions." The other path goes through a module labeled "Group Facts, Annotate Sources, Paragraph Formation," which outputs a "Variation-aware Description." This variation-aware description is then used to generate a "Variations Summary." The diagram uses solid arrows to indicate primary data flow and dashed arrows to indicate optional or supporting input.}, width=\linewidth]{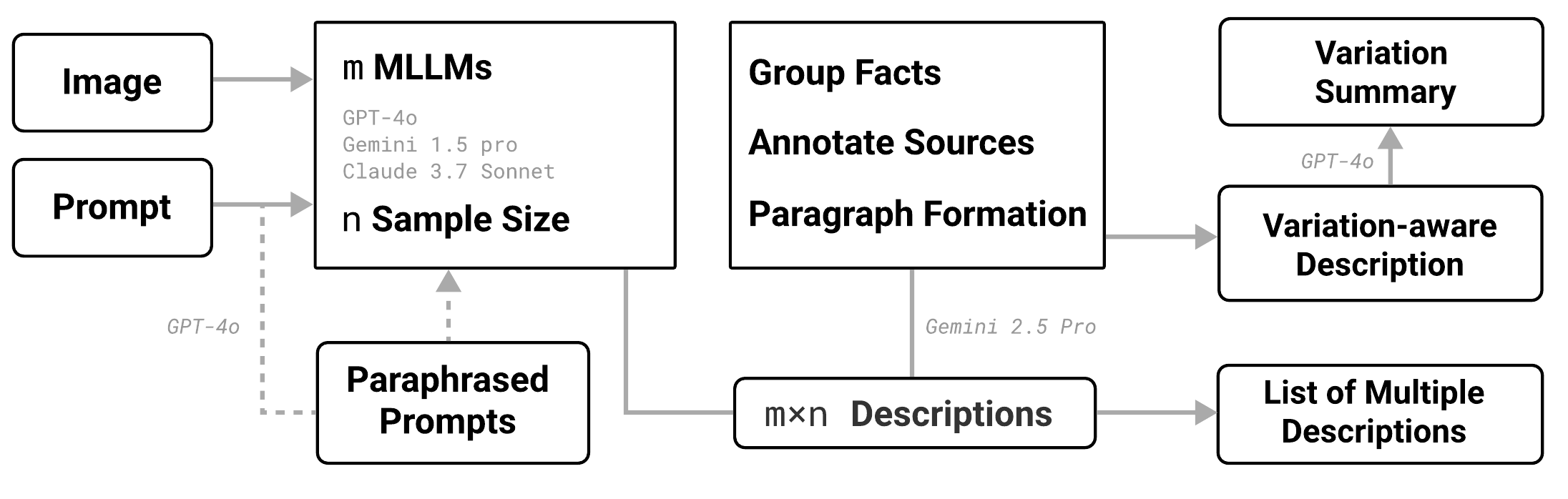}
    \caption{Automatic variation-aware description and variation summary generation pipeline.}
    \label{fig:pipeline}
\end{figure}

\subsubsection{Provenance indicator} Users may want to know what \textbf{trial}, \textbf{prompt}, or \textbf{model} produced each variation. For example, we may say ``the chair is \textit{red} (GPT-4V) or \textit{pink} (Gemini)''. Our prototype includes the model provenance indicator as models may be particularly useful for supporting users assessing model reliability.

\subsubsection{Modality} Our prototype uses \textbf{text} as the primary modality for accessing variations to support efficient screen reader access, while prior work designing variations for sighted people primarily uses text responses coupled with \textbf{visualizations} to show variation. Future work could explore using \textbf{sound} (e.g., lower volume for less supported variants) or even more futuristic modalities such as \textbf{haptics} to support variation assessment. 

\begin{figure*}[htb]
    \centering
    \includegraphics[alt={Four labeled panels (A–D) show how multiple MLLM outputs are aggregated into variation‑aware descriptions. (A) Input photo of a home interior wall with framed art, a chair holding a laundry basket, and a small floral‑covered side table; prompt: “Describe the room setting. Does this wall setting look okay?” (B) Individual model descriptions (Gemini, GPT, Claude). (C) Hierarchical variation‑aware Description groups cross‑model observations: room is a living space (bedroom or den); walls soft green or gray textured; main furniture (bed or armchair or loveseat); other noted items (laundry basket on chair, small side table); wall decor pieces incl. larger artwork and red‑flower print; subjective opinions split between cohesive/cozy vs cluttered. (D) Variation summary. Agreements (living space, small table, framed decor), Disagreements (main furniture type; decor cohesion), and Unique Mentions (GPT mentioned patterned fabric; Gemini mentioned hanging tassel; Claude mentioned traditional and vintage style).}, width=\linewidth]{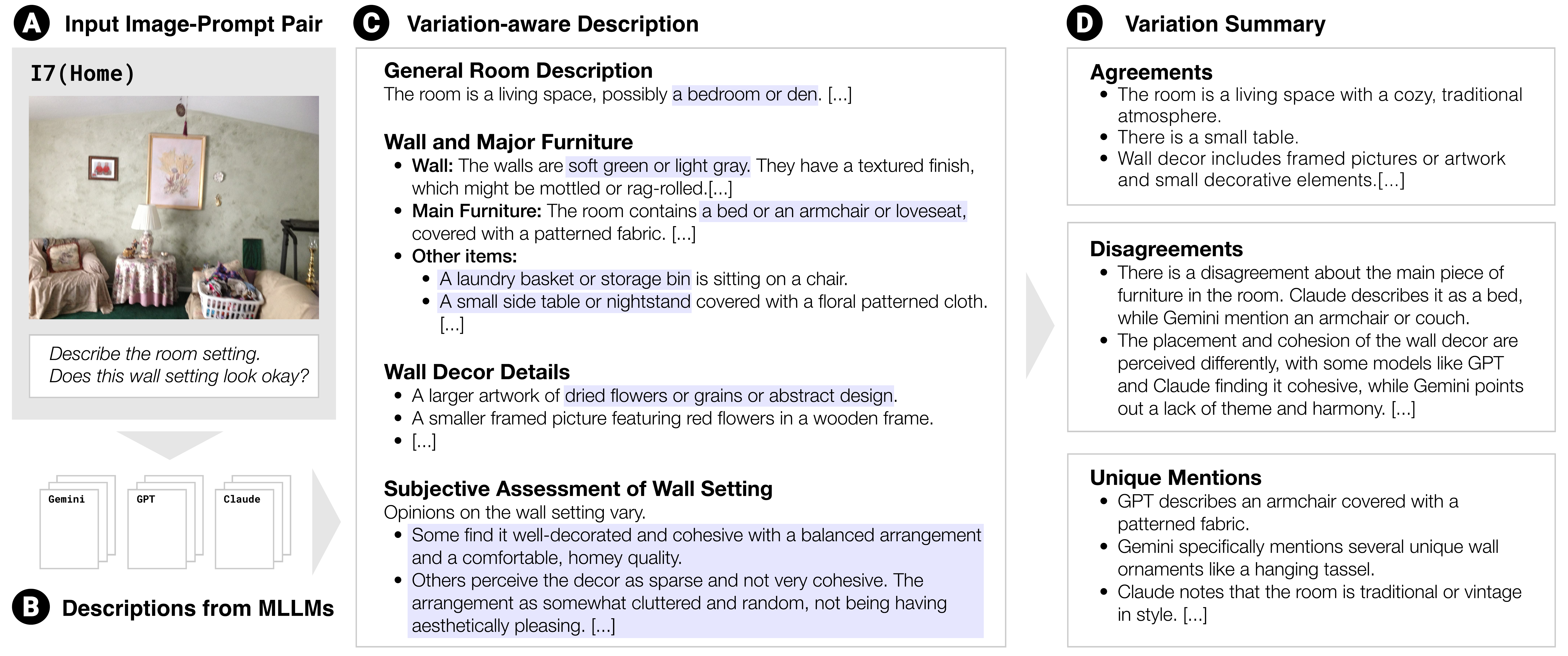}
    \caption{(A) Input image and prompt. (B) Raw image descriptions from 3 MLLMs (\texttt{GPT‑4o}, \texttt{Gemini‑1.5‑Pro}, \texttt{Claude‑3.7‑Sonnet}). (C) Variation‑aware description aggregates all model outputs into a hierarchical markdown. Major variations are highlighted in \materialhl{indigo}. (D) Variation summary further surfaces key agreements, disagreements, and unique mentions across models.
}
    \label{fig:presentation-styles}
\end{figure*}

\subsection{Prototype}
To gather BLV people's perceptions of variations, we develop a study prototype that automatically aligns and presents variations guided by our design choices and accessibility considerations for generative AI systems ~\cite{weisz2024design, mankoff_accessibility_2025}.

\subsubsection{Interface}
Users can upload an image from either their photo albums or a webpage and enter a prompt to query the models on our interface. They may select one or more of the three available models and set query numbers for each model. They can also choose to use either the original prompt or paraphrased versions. For our study, we query each model 3 times per image using the original prompt, yielding a total of 9 descriptions per image. We choose this number because it can harness intra-model and inter-model variations while balancing computational costs and minimizing the cognitive load for participants reading the descriptions. The total number of generated responses also aligns with prior research of similar tasks ~\cite{gero_supporting_2024}.

We show the three variation comparison support styles --- list of multiple descriptions, variation-aware descriptions, and variation summaries (Figure~\ref{fig:presentation-styles}). The \textbf{list} of multiple descriptions displays all original model-generated descriptions in a table for side-by-side comparison. The \textbf{variation-aware description} (Figure~\ref{fig:presentation-styles}C) presents an aggregated detail summary in a hierarchical, coherent markdown format to support a detailed understanding of variations. The \textbf{variation summary} (Figure~\ref{fig:presentation-styles}D) builds on the aggregated summary by explicitly highlighting areas of agreement, disagreement, and unique mentions across models to support quick surface of key variations across all descriptions.

In the variation-aware description, users can toggle how the degree of support for variants is displayed (Table~\ref{tab:support-indicator}). One option shows all variations along with their sources (e.g., \textit{``A small side table (3 of 3 GPT, 2 of 3 Gemini) or a nightstand (1 of 3 Claude)''}), which preserves transparency and attribution. Alternatively, users can view only the relative frequency of agreement (e.g., \textit{``A small side table (56\%) or a nightstand (11\%)''}), natural language indicators (e.g., \textit{``A small side table (moderately supported) or a night stand (poorly supported)''}), or choose to hide this information entirely. 

\subsubsection{Automatic Variation Summary Pipeline}
We select 3 state-of-the-art MLLMs, namely \texttt{GPT-4o}, \texttt{Claude 3.7 Sonnet}, and \texttt{Gemini 1.5 Pro}, to generate image descriptions. Our selection rationale was based on their relevance to visual access technologies and strong visual understanding capabilities. \texttt{GPT-4o} powers the Be My AI application~\cite{BeMyAI}. \texttt{Gemini-1.5-Pro} is the advanced version of the model powering Google's Talkback screen reader and offers limited free access~\cite{Gemini}. \texttt{Claude-3.7-Sonnet} is an another representative recent model with strong reasoning abilities~\cite{noauthor_claude_nodate}. 

We then generate the variation-aware description based on the MLLM-generated image descriptions (Figure ~\ref{fig:pipeline}). We employ Chain-of-Thought ~\cite{wei_chain--thought_2023} and few-shot prompting techniques to instruct \texttt{Gemini-2.5-pro} to decompose each description into atomic facts and reassemble them into coherent, logically structured text. For instance, statements like \textit{``a person is wearing a red shirt''}, \textit{``the individual has on a crimson top''}, and \textit{“someone dressed in red”} all describe the same attribute using different surface forms. Facts that refer to the same aspect but disagree with each other, such as \textit{``the shirt is red''} and \textit{``the shirt is orange''}, are clustered together. Within each cluster, we combine atomic facts about the same subject into single sentences. When variations of a fact exist, we concatenate them using ``or''. These grouped facts are then merged into paragraphs, ensuring that all distinct claims are retained. We annotate differences between model-generated facts and preserve metadata about the source model and original response. We instruct the model to output the summary in markdown format to support BLV users easily navigating the descriptions hierarchically. Finally, we transform the variation-aware description into a more concise variation summary that highlights agreements, disagreements, and unique mentions across models.
\begin{table*}[t]
\centering
\small
\setlength{\tabcolsep}{4pt}
\renewcommand{\arraystretch}{1.2}
\begin{tabularx}{\textwidth}{@{}>{\raggedright\arraybackslash}X
                                >{\raggedright\arraybackslash}X
                                >{\raggedright\arraybackslash}X
                                >{\raggedright\arraybackslash}X@{}}
\toprule
\textbf{None} &
\textbf{Language} &
\textbf{Percentage} &
\textbf{Source} \\
\midrule
There are two \hlg{white} chairs on the left and a grey sofa on the right. At the center there is a white coffee table with a \hlp{marble or glass or wood} top and a gold base. There is a built-in shelf on the back wall with decorative items, like \hlb{books and a television}. 
&
There are two \hlg{white (well-supported)} chairs on the left and a grey sofa on the right. At the center there is a white coffee table with a \hlp{marble (moderately supported) or glass (poorly supported) or wood (very little support)} top and a gold base. There is a built-in shelf on the back wall with decorative items, like \hlb{books (moderately supported) and a television (moderately supported)}. 
&
There are two \hlg{white (100\%)} chairs on the left and a grey sofa on the right. At the center there is a white coffee table with a \hlp{marble (56\%) or glass (33\%) or wood (11\%)} top and a gold base. There is a built-in shelf on the back wall with decorative items, like \hlb{books (33\%) and a television (33\%)}. 
&
There are two \hlg{white (3 of 3 GPT, 3 of 3 Gemini, 3 of 3 Claude)} chairs on the left and a grey sofa on the right. At the center there is a white coffee table with a \hlp{marble (3 of 3 GPT, 2 of 3 Gemini) or glass (3 of 3 Claude) or wood (1 of 3 Gemini)} top and a gold base. There is a built-in shelf on the back wall with decorative items, like \hlb{books (3 of 3 GPT) and a television (3 of 3 Gemini)}. \\
\bottomrule
\end{tabularx}
\caption{Variation-aware description without support indicator and with three variant support indicators we designed (Language, Percentage, Source). Agreements (top, \hlg{green}), disagreements (middle, \hlp{red}), and unique mentions (bottom, \hlb{blue}) are highlighted.}
\label{tab:support-indicator}
\end{table*}

\subsubsection{Implementation}
The prototype's frontend is built using React, and the backend runs on a Python Flask server. We followed the guidelines of W3C \cite{wai} and tested the compatibility with all three major screen readers: NVDA, JAWS, and VoiceOver. Major prompts used in the study are provided in the Appendix.

\section{User Study}
We conducted a within-subject study with \numpar{} BLV participants to investigate how surfacing variations in MLLM-generated image descriptions impacts BLV users' ability to recognize unreliable claims in MLLM responses and their perceived reliability of MLLM responses. 

\begin{itemize} 
\item[\textbf{RQ1:}] What is the impact of surfacing variations on BLV users' ability to recognize unreliable claims and their percieved reliability of MLLM responses?
\item[\textbf{RQ2:}] What are effective design strategies for surfacing variations in MLLM-generated image descriptions for screen reader users?
\item[\textbf{RQ3:}] What are the potential use cases, benefits, and limitations of surfacing variations in image descriptions for BLV users? 
\end{itemize}

\subsection{Participants}
\begin{table*}[t]
\centering
\resizebox{\linewidth}{!}{%
\begin{tabular}{cllllll}
\toprule
\textbf{PID} & \textbf{Age} & \textbf{Gender} & \textbf{Visual Impairment} & \textbf{Onset} & \textbf{Screen Reader(s)} & \textbf{Prior MLLM-Powered Tool Use} \\ \midrule
1 & 41 & M & Totally blind & Birth & JAWS & Be My AI, ChatGPT \\
2 & 24 & F & Light perception & Birth & VoiceOver & Be My AI, PiccyBot, Seeing AI \\
3 & 33 & M & Totally blind & Birth & JAWS, NVDA, Chrome, Talkback, VoiceOver & AccessAI, Be My AI, ChatGPT, Claude, PiccyBot \\
4 & 32 & M & Totally blind & Birth & JAWS, VoiceOver & Be My AI \\
5 & 33 & M & Totally blind & 17 & VoiceOver & Be My AI, Seeing AI \\
6 & 30 & F & Light perception & Birth & TalkBack, VoiceOver, NVDA, JAWS & Be My AI, OOrion, Seeing AI \\
7 & 22 & M & Totally blind & Birth & VoiceOver & AccessAI, Be My AI, ChatGPT, Claude, Gemini, Grok \\
8 & 39 & F & Light Perception & born & VoiceOver & AccessAI, Be My AI, ChatGPT, Meta Rayban, Seeing AI \\
9 & 46 & F & Totally blind & 2 & JAWS, NVDA & Be My AI, ChatGPT \\
10 & 30 & M & Totally blind & 5 & JAWS, NVDA, VoiceOver & AccessAI, Be My AI, ChatGPT, Gemini, Maestro, Seeing AI \\
11 & 29 & F & Light perception & Birth & JAWS, NVDA, VoiceOver & Be My AI, ChatGPT, EnvisionA11y \\
12 & 57 & F & Light perception & Birth & JAWS, NVDA, VoiceOver, Narrator & AccessAI, Be My AI, ChatGPT, Gemini, Picture Smart w/ JAWS \\
13 & 35 & F & Light perception & Birth & JAWS, NVDA, VoiceOver, Narrator & AccessAI, Be My AI, ChatGPT, PiccyBot, Seeing AI \\
14 & 50 & M & Light perception & 9 & JAWS, VoiceOver, NVDA & AccessAI, Be My AI, ChatGPT, Gemini, NotebookLM \\
15 & 55 & F & Totally blind & Birth & JAWS, NVDA, VoiceOver, Talkback & Vision AI Assistant \\
\bottomrule
\end{tabular}
}
\caption{Participant details for BLV participants in the user study including their participant ID, age, gender self-described visual impairment, age of onset, and prior use of screen reader(s) and MLLM-powered tools.}
\label{tab:participant_details}
\end{table*}
We recruited \numpar{} BLV participants who regularly use screen readers to access online content and have prior experience with MLLM-powered visual access tools for image descriptions (Table~\ref{tab:participant_details}). Participants were recruited through BLV community mailing lists. They reported using a variety of screen readers (e.g., NVDA, TalkBack, JAWS, and VoiceOver) and have rich experiences with diverse MLLM-powered tools (e.g., Be My AI, AccessAI, and PiccyBot). They use these tools across diverse scenarios, including interpreting images in messages and on social media, selecting outfits, reading text and numbers in books and professional documents. Among our participants, 9 were totally blind, and 6 had some degree of light perception. Participant ranged from 22 to 57 years old.

We asked participants about their past experiences with AI-generated image descriptions. They use these tools across diverse scenarios, including interpreting images in messages and on social media, selecting outfits, and reading text and numbers in books and professional documents. On a 7-point scale (1 = not reliable at all, 7 = very reliable), participants rated the overall reliability of MLLM-generated image descriptions at 4.56 on average (SD = 1.09). While many participants appreciated the convenience these tools offer, several emphasized the importance of fact-checking in high-stakes situations. Participants with more experience using MLLM-powered tools were also more aware of common error patterns. For example, P2 noted, \textit{``Be My AI was really bad at numbers in the beginning, but now it seems to be much better.''} Participants fact-check by retaking photos, using multiple tools for comparison, and asking others for verification, which aligns with findings from prior work ~\cite{alharbi2024misfitting}.

\subsection{Materials}
\begin{figure}[t]
    \centering
    \includegraphics[alt={A 3×3 image grid displays nine study images categorized by ambiguity source type. The first row, labeled "Model Limitations," includes: I1 (Map), I2 (Swiftie), and I3 (Washing Machine). The second row, "Image Quality," includes: I4 (Screen), I5 (Medication), and I6 (Card). The third row, "Subjectivity," includes: I7 (Home), I8 (Outfit), and I9 (Dragonfly). Each image is labeled with an ID and a brief descriptor.}, width=3.33in]{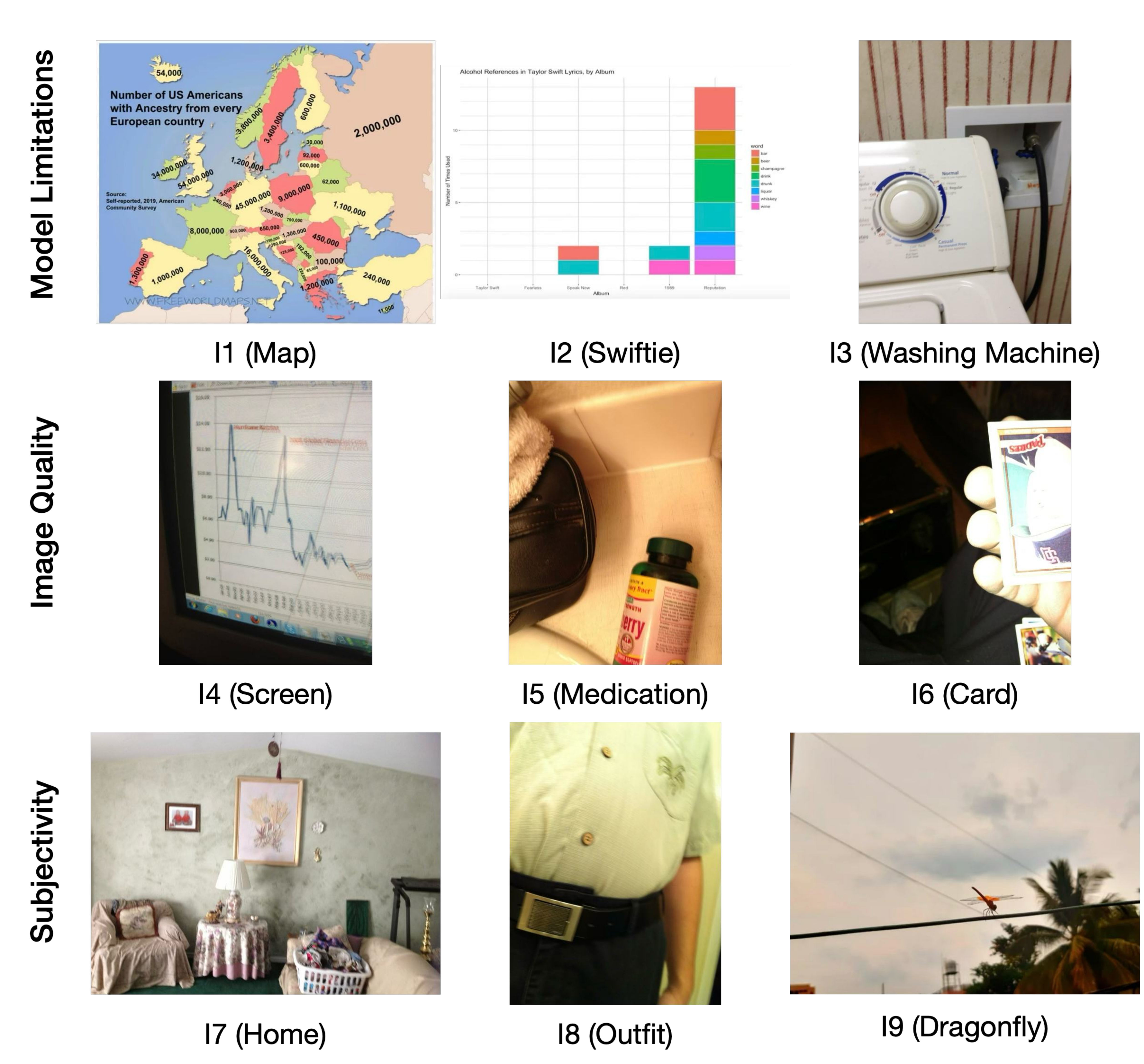}
    \caption{Images used in the study with corresponding name, label, and category.}
    \label{fig:images}
\end{figure}

We selected 9 challenging image–prompt pairs that include ambiguity and reflect common use scenarios for BLV MLLM users (Figure~\ref{fig:images}). 6 images were selected from the VizWiz dataset, and 3 were selected from public threads on Reddit. The VizWiz dataset contains images and questions collected directly from BLV users, making representative real-world use cases~\cite{vizwiz}. However, this dataset is relatively dated and lacks coverage of diagrams ~\cite{zhao_tada_2024} and social media content that BLV users increasingly engage with ~\cite{bennett2018teens}. To address this gap, we selected 3 images from popular Reddit thread posts that contain minimal accompanying text. Details of these images can be found in Table~\ref{tab:images_selected}. We categorized the selected images based on three primary sources of ambiguity that are likely to induce variation in MLLM-generated image descriptions: \\ 
% \amy{here's fine for now can move later}\meng{maybe move this part to section 3}

\noindent\textbf{Model limitation:} MLLMs still have imperfect performance for certain tasks such as spatial reasoning and thus produce misinterpretations of relative positions, shapes, or layouts in structured visuals like maps and graphs ~\cite{li-etal-2023-evaluating, liu2024survey}.\\ 

\noindent\textbf{Image quality:} Images taken by blind people are often poorly cropped, blurry, or incorrectly oriented~\cite{gurari2018vizwizgrandchallengeanswering}. Even sighted users may struggle to interpret such images, and models similarly may hallucinate or make incorrect inferences ~\cite{10.1145/2384916.2384941, huh2024long} for unclear images. \\ 

\noindent\textbf{Subjectivity:} Tasks that involve evaluating fashion, room aesthetics, or pet appearance introduce ambiguity, as humans may disagree ~\cite{bennett2018teens}.

\subsection{Task}
In this study, participants were asked to identify potentially unreliable parts of MLLM-generated image descriptions. We defined a part of the description as unreliable if it fell into one of the following categories: (1) \textbf{incorrect}, or the contains information that is likely false, (2) \textbf{speculative}, or the claim includes information that cannot be verified from the image alone, and (3) \textbf{opinionated}, or the claim reflects a subjective opinion.
% These categories are adapted from our previous ambiguity categories that cause variations. 
Each image was paired with a standardized prompt structure: ``Describe [...]'' followed by a targeted visual reasoning question (e.g., \textit{``Describe this object. What is inside the container?''}) to ensure MLLM-generated descriptions were rich in content and aligned with task-specific goals. For each image, participants were first presented with a hypothetical scenario (e.g., \textit{``You found a bottle, but you are not sure what it is.''}) along with the exact prompt used to generate the description.

Participants were then shown one of three description conditions:
\begin{enumerate}
    \item \textbf{Single:} A single MLLM-generated description;
    \item \textbf{List:} A list of 9 descriptions generated from 3 trials of 3 MLLMs;
    \item \textbf{Ours:} Variation-aware description and variation summary generated based on the same 9 descriptions.
\end{enumerate}

For each image, participants had up to 4 minutes to verbally report any unreliable parts of the description(s) and explain their reasoning. Afterward, they answered questions about their perception of the description's quality ~\cite{levinboim2019quality}, information coverage ~\cite{levinboim2019quality}, and trustworthiness ~\cite{gardner2020determining}. 

% Questionnaire used in the study can be found in Appendix.

\subsection{Procedure}
\noindent\textit{Study setup and pre-study questionnaire (10 minutes).} Each study session lasted 1.5 hours and was conducted one-on-one over Zoom. The study protocol was approved by our institution’s IRB, and participants received \$30 USD per hour as compensation for their time. The study began with a pre-study questionnaire on demographic information, experience in MLLM-powered visual access tools, and scenarios where they are using these tools. We also asked participants about their current level of trust in MLLM-generated image descriptions and when and how they fact-check such descriptions.  \\ 

\noindent\textit{Task (60 minutes).} 
We shared a secure link to our web-based task interface with each participant. At the beginning of the session, we walked participants through a 5-minute tutorial to demonstrate the interface and explain the different presentation styles. Participants then evaluated the reliability of image descriptions generated for each of our 9 pre-selected images, consisting of 3 images for each of the 3 ambiguity sources (model limitation, image quality, and subjectivity). For each image, participants used one of three description conditions (single, list, ours), and we counterbalanced description conditions across the three ambiguity sources such that participants used each description condition 3 times, once for each ambiguity source. We randomized the sequence of images for each participant. We randomized the image sequence for every participant. For each image, participants had up to 4 minutes to read the descriptions and complete a Likert-scale questionnaire.\\

\begin{figure*}[t]
    \centering
    \includegraphics[alt={Two bar charts compare three presentation styles—Single (red), List (yellow), and Ours (blue)—on two metrics: the number of identified unreliable claims (left) and perceived reliability (right). The left chart shows that "Ours" led to significantly more unreliable claims being identified across all categories (Overall, Model Limitation, Image Quality, and Subjectivity). The right chart shows ``Single'' descriptions were rated highest in perceived reliability, followed by ``List'', with ``Ours'' rated lowest in several categories. Statistical significance is indicated with asterisks (* p < 0.05, ** p < 0.01, *** p < 0.001).},width=\linewidth]{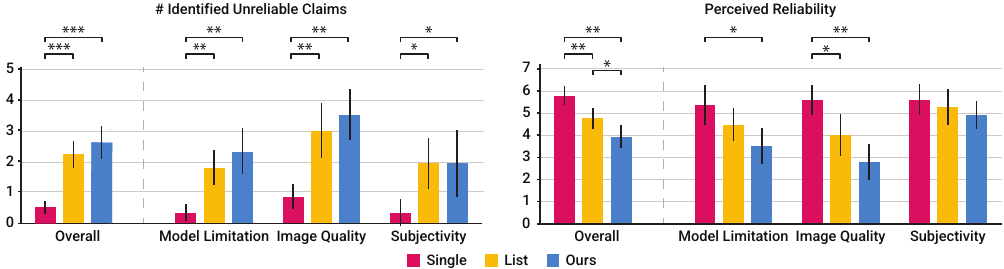}
    \caption{Left: average identified unreliable claims reported by participants overall and in each image category. Right: average perceived reliability rating (1 = not reliable at all, 7 = most reliable) overall and in each image category. Error bars represent a 95\% confidence interval. We applied the Friedman test followed by pairwise Wilcoxon signed-rank tests with Bonferroni correction. Significance is marked as * $p < 0.05$, ** $p < 0.01$, and *** $p < 0.001$.}
    \label{fig:results-bars}
\end{figure*}

\noindent\textit{Open-ended system use and semi-structured interview (20 minutes).} After the task, participants uploaded up to 2 images of their choice to try our interface on the scenarios that they think variations will be useful. They freely explored different presentation styles and provided open-ended feedback.  To conclude the session, we conducted a 10-minute semi-structured interview. Participants reflected on the strengths and limitations of variation-aware descriptions, shared their thoughts on the interface designs, and discussed how such tools could support their everyday lives.\\

\subsection{Analysis}
We adopted a mixed‑methods approach to analyze participants’ responses. Two researchers categorized the participant-identified unreliable claims into one of the three categories from Section 4.3 independently and met to resolve discrepancies. We also categorized incorrect and speculative claims as either a true positive (correctly identified) or a false positive (incorrectly identified) by examining the claim and image together (Table~\ref {tab:unreliable-claim-stats}). Finally, we categorized comments that did not fit into any categories as irrelevant (e.g. \textit{``I feel like something’s missing''}).  We separately analyzed transcribed comments for identifying unreliable claims (during the task) and from the semi-structured interview (after the task). For both, two researchers independently performed open coding on the transcripts and observation notes, clustering codes with affinity diagrams. For the former, we identified major reasons for identifying unreliable claims (Table~\ref {tab:single-stats}). For the latter, our codes mirrored the interview questions (e.g. benefits of variations). For the quantitative analysis, one researcher counted and categorized the number of identified unreliable claims across the three description conditions (Single, List, Ours). The same researcher extracted Likert‑scale ratings of perceived reliability, computed descriptive statistics, and performed Friedman tests with post‑hoc Wilcoxon signed‑rank comparisons to check condition effects.

\section{Results}
Overall, \textbf{surfacing variations in MLLM responses increased the number of unreliable claims identified (RQ1)} in MLLM descriptions by 4.9x for \textbf{ours} (M = 2.62 , SD = 1.72) or 4.2x for \textbf{list} (M = 2.24, SD = 1.52) compared to presenting a \textbf{single} description (M = 0.53, SD = 0.73) (Tabe ~\ref{tab:unreliable-claim-stats}). 
A Friedman's test indicated a significant impact of condition on number of unreliable claims overall ($p < 0.001$) with pairwise comparisons indicating a significant difference between ours and single ($p < 0.001$), and list and single ($p < 0.01$) (Figure~\ref{fig:results-bars}).

\textbf{Surfacing variations in MLLM responses also decreased the perceived reliability of MLLM responses (RQ1)} from 5.78 (SD = 1.41) of 7 for a single description to 4.76 (SD = 1.61) of 7 for a list of descriptions and 3.93 (SD = 1.70) for ours. 
% (Table ~\ref{tab:perceived-reliability}). 
A Friedman's test indicated a significant impact of condition on perceived reliability of MLLM-generated image descriptions overall ($p < 0.01$) with pairwise comparisons indicating a significant difference between all pairs (Figure~\ref{fig:results-bars}).

11 of 15 participants ranked our variation summary (ours) as their favorite option with 9 of 15 participants ranking variation-aware descriptions (ours) as their second favorite option, while only 5 participants rated the list of descriptions (list) or a single description as their first or second favorite indicating strong support for our new aggregated variation approaches (\textbf{RQ2}) (Figure~\ref{fig:pref-presentation}). All BLV participants wanted to use our variation surfacing prototype in the future for a variety of purposes from \textbf{high-stakes scenarios} such as assessing the path of an incoming tornado (P12) to \textbf{obtaining subjective critiques} for social media posts (P15) (\textbf{RQ3}).

In the rest of this section, we share findings on how users identified unreliable information with and without variations (\textbf{RQ1}), the benefits and drawbacks of different display options in our prototype sampled from our design space (\textbf{RQ2}), and when surfacing variations in MLLM descriptions mattered for BLV participants (\textbf{RQ3}). 

\begin{table*}[t]
\centering
\small
\begin{tabular}{@{}l ccc ccc ccc ccc@{}}
\toprule
 & \multicolumn{3}{c}{\textbf{Model Limitation}} & \multicolumn{3}{c}{\textbf{Image Quality}} & \multicolumn{3}{c}{\textbf{Subjective}} & \multicolumn{3}{c}{\textbf{Overall}} \\ 
\midrule
 & \textbf{Single} & \textbf{List} & \textbf{Ours} & \textbf{Single} & \textbf{List} & \textbf{Ours} & \textbf{Single} & \textbf{List} & \textbf{Ours} & \textbf{Single} & \textbf{List} & \textbf{Ours} \\ 
\midrule
\textbf{True Positive (Incorrect)}   & {\ul 4}  & 18 & \textbf{29} & {\ul 3}  & 20 & \textbf{27} & {\ul{0}} & 4 & \textbf{7}  & {\ul 7}  & 42 & \textbf{63} \\
\textbf{True Positive (Speculative)} & 0        & 0  & 0           & {\ul 8}  & 19 & \textbf{22} & {\ul 3} & \textbf{8} & 6 & {\ul 11} & 27 & \textbf{28} \\
\textbf{False Positive}              & {\ul 1}        & \textbf{6} & 5 & {\ul 3} & \textbf{7} & 4 & {\ul 1} & \textbf{4} & \textbf{4} & {\ul 5} & \textbf{17} & 13 \\
\textbf{Opinionated}                 & {\ul 0}  & 1  & \textbf{2}  & 0 & 0 & 0 & {\ul 1} & \textbf{11} & \textbf{11} & {\ul 1} & 12 & \textbf{13} \\
\textbf{Irrelevant}                  & {\ul 0}        & \textbf{1} & {\ul 0} & {\ul 0} & \textbf{1} & {\ul 0} & {\ul 0} & \textbf{1} & \textbf{1} & {\ul 0} & \textbf{3} & 1 \\ 
\midrule
\textbf{Total Count} & {\ul 5} & 26 & \textbf{36} & {\ul 14} & 47 & \textbf{53} & {\ul 5} & 28 & \textbf{29} & {\ul 24} & 101 & \textbf{118} \\
\textbf{Mean}       & 0.33 & 1.73 & 2.40 & 0.93 & 3.13 & 3.53 & 0.33 & 1.87 & 1.93 & 0.53 & 2.24 & 2.62 \\
\textbf{Std. Dev.}  & 0.49 & 1.03 & 1.30 & 0.70 & 1.55 & 1.51 & 0.82 & 1.60 & 1.98 & 0.73 & 1.52 & 1.72 \\
\bottomrule
\end{tabular}
\caption{Total count, means, and standard deviations of identified unreliable claims under three conditions. Counts are further broken down by image category (Model Limitation, Image Quality, Subjectivity) and further classified as True Positive (Incorrect and Speculative), False Positive, Opinionated, or Irrelevant. The highest value in each group is in \textbf{bold}. The lowest is  \underline{underlined}.}
\label{tab:unreliable-claim-stats}
\end{table*}

\subsection{What Makes Image Descriptions Appear Unreliable?}
In our analysis of participants explanations for identifying claims as ``unreliable,'' participants predominantly take \textbf{inconsistency} (96\% of all reported claims in List condition; 94\% of all reported claims in Ours condition) as the indicator of unreliable claims when multiple descriptions are available, while they take \textbf{lack of details} as the major indicator (54\% of all reported claims) when they assess a single description. 

\begin{table}[]
\begin{tabular}{@{}ll@{}}
\toprule
\textbf{Cause of Unreliability in Single Description} & \textbf{Count} \\ \midrule
\textbf{Lack of Details}          & 13             \\
\textbf{Suggestive and Uncertain Language}     & 5              \\
\textbf{Prior Experience with MLLMs}          & 4             \\
\textbf{Others}                   & 2              \\
\midrule
\textbf{Total}                   & 24             \\ \bottomrule
\end{tabular}
\caption{Breakdown of the reasons for each unreliable claim reported in Single condition.}
\label{tab:single-stats}
\end{table}

\subsubsection{Identifying unreliable claims with single descriptions.}
\textbf{Lack of details} is the most popular indicator. Participants frequently flagged descriptions that were too general or lacked reliable critical details. 6 participants (P1, P4, P7, P9, P10, P11) expressed frustration that language models often left out essential elements during the study. P11 said description for I6 (Medication) omitted dosage and ingredient details, but these are \textit{``actually what I want''}. Similarly, P9 felt confused when reading the description for I4 (Screen), \textit{``I know it is a graph, but why doesn't it tell me anything about the x-axis?''} P9 and P10 both consider length and detail information as major source to evaluate reliability. \textit{``If one description is picking up more thoroughly, the second is very sketchy, the third one is in the middle. I would not use the second one because it is sketchy and does not give me any help.''} (P10) 5 of 24 unreliable claims were flagged due to \textbf{suggestive and uncertain language} (Table ~\ref{tab:single-stats}). BLV participants were sensitive to phrases that indicated speculation or guesswork. \textit{ ``The suggestive language makes me think...is this reliable?''} (P6) Likewise, P12 criticized vague phrasing in the description of I6 (Card): \textit{``It appears to be baseball trading card… This description later says 'from the Phillies baseball team,' but why does it only use 'appears to be a baseball card'? Why?''} On the contrary, P6 thought the description for I2 (Swiftie) is very reliable because \textit{`` it didn’t seem to give suggestive language. It didn’t say `perhaps' or `there may be', it sounded more factual''}, yet the image contains one of the most factual errors. Participants also drew on \textbf{prior experience with MLLMs} to judge reliability. P3 and P10, who have relatively more experience in using MLLM-powered tools, were aware that models often struggle with numbers and choose to distrust numbers mentioned in image descriptions of I1 (Map). \\

\subsubsection{Identifying unreliable claims with variations.} 
Participants naturally looked for differences to assess reliability when multiple descriptions are presented. The most common strategy was to look for \textbf{inconsistencies between descriptions}. Participants utilized differences as a cue to identify 97 among 101 reported claims under List condition and 112 among 118 reported under Ours condition even when suggestive language and lack of specificity still exist. Many participants used the degree of agreement across models as a signal for how trustworthy a specific claim might be. As P5 explained, \textit{``If the difference is trivial, then it should be OK, but the number of US ancestry from the UK jumps too much.''} When variations spilt a lot, they also think it is less reliable, P12 pointed out, \textit{``We don’t have any strong percentages here so I wouldn't think this is reliable.''} Participants were also often cautious about claims that appeared in only one or two descriptions. When a detail was uniquely mentioned, it raised concerns about its reliability. \textit{``Only one [description] mentioned x-axis and y-axis in the image. It is what I need, but I'm not sure if it is correct.''} (P10) However, not all unique mentions are necessarily errors. Some response variations simply reflect the model’s interpretation or attention, but participants may deem them as potentially incorrect information for different parts of the image. In complex images like I1, for example, there was so much information on the map that descriptions only pick up on some numbers that other models do not. P11 noted, \textit{``Romania was seen in some models, but not in other models.''} and flagged it as an unreliable part even when the claim was correct. 

\subsection{What are Effective Design Strategies for Surfacing Variations?}

Participants ranked the variation summary as their favorite presentation style (11 of 15) and the variation-aware description as their second favorite one (9 of 15) because they can help them quickly find the unreliable claims in the image descriptions. Support indicator is helpful, but it largely depends on personal preferences. Most participants (14 of 15) believed text is already effective in conveying variations, but open to alternative modalities beyond text. 7 of 15 participants were enthusiastic about having the ability to switch between presentation styles based on different contexts. \textit{``It depends on what kind of information I want from the image. I think models are generally accurate on scene description so I would just like a list to freely surface the differences. But if the image is quantitative like graphs, I would want the summary.''} (P7) With images for which the prompt is more subjective, \textit{``having them all grouped together like that at one time is helpful [...] I would with people getting different viewpoints, different perspectives from the models, and normally I wouldn't have that.''} (P14).

\begin{figure}[htb!]
    \centering
    \includegraphics[alt={A horizontal stacked bar chart titled "Preference on Presentation Style" compares user rankings (1 = most preferred, 4 = least preferred) across four styles: Variation Summary, Variation-Aware Description, List of Multiple Descriptions, and Single Description. "Variation Summary" received the most Rank 1 votes (11), followed by "Variation-Aware Description" (9). "Single Description" received the most Rank 4 votes (7), indicating it was least preferred overall. Each bar is color-coded by rank.}, width=3.33in]{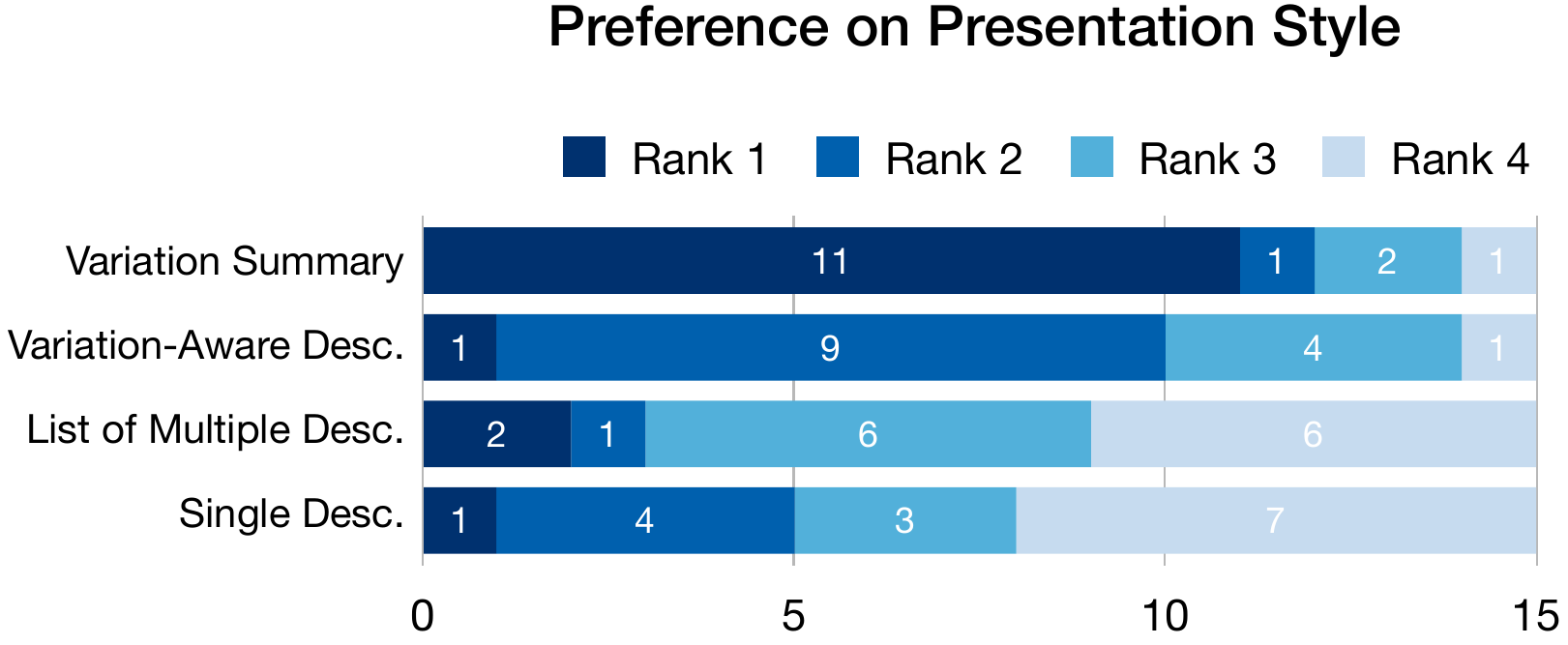}
    \caption{Participants' preference on variation presentation style and single description. (1 = most preferred, 4 = least preferred)}
    \label{fig:pref-presentation}
\end{figure}
\subsubsection{Presentation style}

Participants ranked variation summary (Ours) the highest (M = 1.53, SD = 0.99), followed by variation-aware description (Ours) (M = 2.33, SD = 0.72), list of multiple descriptions (M = 3.07, SD = 1.03). Single description baseline was the least preferred (M = 3.07, SD = 1.03). (Figure ~\ref{fig:pref-presentation}). A Friedman’s test followed by a pairwise Wilcoxon signed-rank test with Bonferroni correction indicated a significant preference for variation summary to single description ($p < 0.05$). 
\\

\noindent\textbf{\textit{Variation Summary:}} 
11 participants (P1-7, P9-11, P15) favor variation summary the most because it is concise and clear, and helps them \textit{``see to what extent and how different AIs are describing them''} (P2) without \textit{``having to go through each individual description''}(P6). P6 reflected that the agreement summary helped her quickly understand where all the models reach a consensus, so she \textit{``feels pretty secure about that because it is a high confidence level''}. The disagreements summary raises their awareness of the conflicting information between multiple descriptions. P10 says that \textit{``you can look at the disagreements, check if those details don't matter or that's super relevant steers to whether you need clarification or not.''} I2 (Swiftie) is a stacked bar graph of the number of alcohol references Taylor Swift made in each of her albums. P11 noticed that the agreement summary tells her the number of alcohol in \textit{Reputation} was the highest, but models disagreed when it came to the references, so she then carefully examined the reference information.
\\

\noindent\textbf{\textit{Variation-aware Description:}} 7 of 11 participants who ranked variation summary the first ranked variation-aware descriptions second (P1, P3-5, P9-10, P15) because they can learn more specific information about the image in depth when they find it necessary. 
P12 states \textit{``it was really nice to have all that detail, especially if it's something that I'm not familiar with, but I'm trying to get a deeper sense and having different perspectives or different verbiage to kind of help build that imagery of what I'm looking at.''} P10, who is a neuroscience PhD student, regularly interprets figures and graphs. He believes that \textit{``variation-aware summaries are good when I am looking at graphs when I need really specific pieces of information.''} 

After encountering variations between multiple descriptions, participants appreciated the additional detail. \textit{``Before this study, I thought, `maybe I would just like something short and sweet all the time.' And then I realized, `no, no, that's too short.'''}(P15) While many participants love the comprehensiveness of variation-aware description, others (P2, P6, P11) feel overwhelmed when going through all of the details. P11 felt that \textit{``the language in there might get a little distracting just trying to keep track of it in your head.'' }However, they also acknowledge that they can scrutinize the information better with this presentation style.
\\

\noindent\textbf{\textit{List of Multiple Descriptions:}} 2 participants (P13, P14) ranked the list presentation style first because they can quickly surface themselves without the need to analyze the summary themselves. P14 appreciated that the information was organized, and having the ability to apply filters is great. \textit{``I would probably do one model at a time rather than multiple models at once.''} when reading the descriptions. However, the rest 13 participants unanimously pointed out that there was too much information to process in the list presentation style, especially for screen readers. Participants have to read the list format and keep track of the consistencies and inconsistencies as they go which creates doubt if they cannot recall all of the information (P4, P5, P7, P11). As a result, participants expressed not wanting to use the list format for simple visual questions, with P7 stating that \textit{``AI descriptions are long. List is way too much because you have to remember all of them. Do I really want to judge an SAT test style graph using a list?''} 
% When multiple descriptions agree on most details in the image, the list format \textit{``can be redundant''} (P15).
\\

\subsubsection{Support Indicator} Participants showed diverse preferences for support indicators. 5 participants (P3, P5, P11, P14, P15) ranked the \textbf{model source} (M = 1.93, SD = 0.88) as their favorite because it is transparent on where models split and can have a better sense of the capability of each model. \textit{``I like it because I know why everything is supported.''} (P5). Yet, they do acknowledge that \textit{``you hear it all the time and you need to keep a tab of the different disagreements or agreements''}(P11). 4 participants (P7, P10, P12, P13) ranked \textbf{percentage} at the top (M = 2.13, SD = 0.99) because it is intuitive and \textit{``screen reader can handles the percentage indicator better''}(P12). 4 participants (P1, P2, P4, P9) preferred \textbf{no support indicator} to any of three (M = 2.80, SD = 1.21) because it is the most \textit{``fluent''} and \textit{``natural''}. Only 2 participants (P6 and P8) ranked \textbf{language} as their first choice (M = 3.13, SD = 1.06) because it was more natural for them compared to numbers, but other participants deemed language as \textit{``ambiguous''} (P1, P3, P11, P12) and \textit{``confusing''}(P7).

\subsubsection{Modality}
One participant (P2) explicitly expressed interest in representing the reliability level in other modalities while other participants (13 of 15) were neutral but open to other modalities because they believed that text can already effectively communicate variation information as it is \textit{``sufficient''} (P14) and \textit{``easy to understand''} (P3). P13, who disliked modalities other than text, shared her struggle with interpreting the darkness sonification feature in Seeing AI and strongly preferred text.

\begin{figure}[htb!]
    \centering
    \includegraphics[alt={A horizontal stacked bar chart titled "Preference on Support Indicator" compares user rankings (1 = most preferred, 4 = least preferred) for four styles: Source, Percentage, None, and Language. "Source" received the most Rank 1 and Rank 2 votes combined (5 and 7), indicating strong preference. "Language" received the most Rank 4 votes (7), making it the least preferred overall. Each bar is divided into four color-coded segments representing ranks 1 through 4.},width=.9\linewidth]{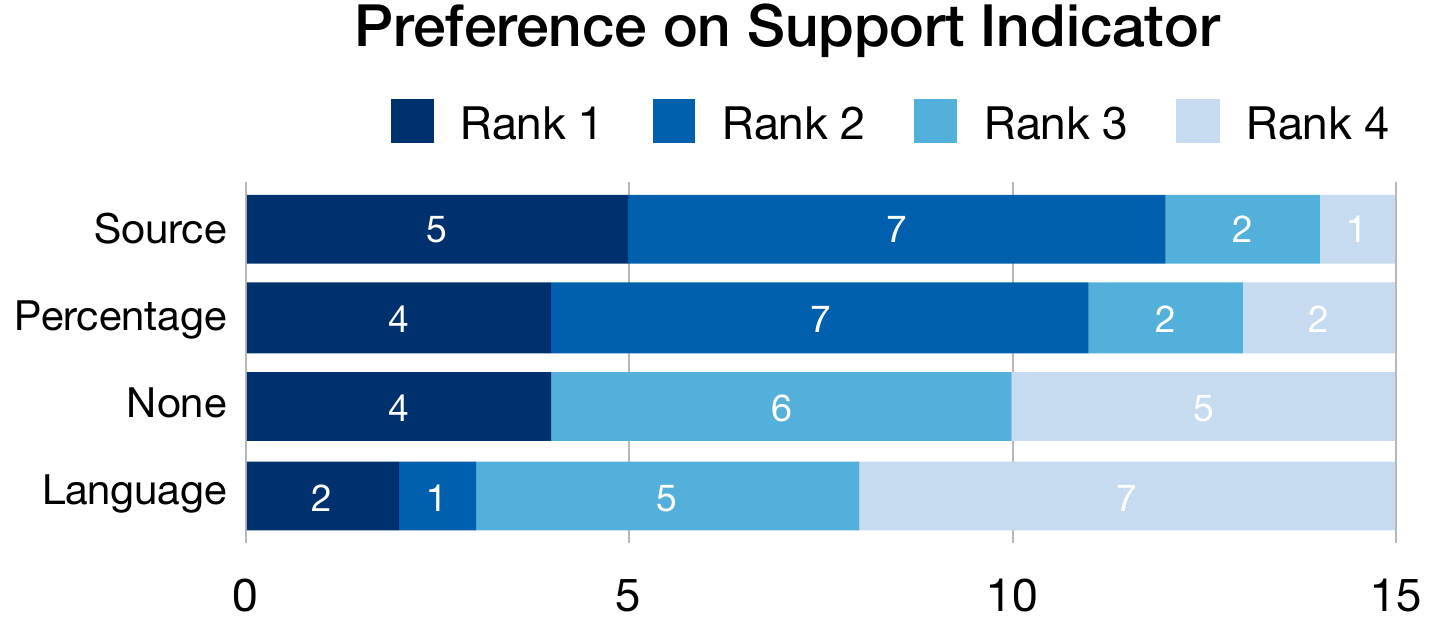}
    \caption{Participants' preference on support indicator styles. (1 = most preferred, 4 = least preferred)}
    \label{fig:pref-indicator}
\end{figure}

\subsection{ When do Variations in MLLM-generated Image Description Matter? }

Participants found variations most useful in \textbf{high-stakes} and \textbf{subjective} scenarios. All participants expressed strong interest in reading variation-aware descriptions in detail when the consequences of inaccuracies are serious. 14 of the 18 images that participants selected in the open-ended system use are high-stakes, including healthcare (P1, P2-1, P5, P10), appliance usage (P3), navigation (P6), stock price (P7-1), shopping (P2-2, P7-2, P9-1, P9-2, P11), tornado path (P12), comic book (P14-1), and accessibility test (P14-2) (Table ~\ref{table:open-end-use}).
Participants also found variations helpful for their own fact-checking scenarios. Three participants see variation-aware descriptions as best applicable for outfits (P6, P11, P2) or makeup (P12). P8 also briefly mentioned using these descriptions for reading insurance documents. P6 uniquely adds that she would find it helpful in \textit{``her day-to-day life with street signs.''} P13 also stated finding this tool specifically applicable in her own life, doing \textit{``workbook work with [her] child.''}

Six of 15 (P6, P7, P8, P10, P13, P14) participants noticed the subjective discrepancy between different models and would love to understand the differences in the future. 2 of the 18 images that participants selected in the open-ended system use are looking for creative interpretation of the comic book (P14) and critique of the photo they took (P15). When reading through the descriptions for I9 (Dragonfly), P14 was able to get a general idea of the image content, so having different opinions from different models is a plus compared to just relying on one model. P10 frequently posts pictures about his life and his guide dog on social media, but he is always concerned about whether the picture is appropriate. \textit{``I would use this tool to describe my photos for social media posts. Right now, I take a lot of images and then upload them to ChatGPT to tell me which one is the best. This system tells me that one of the models says that the dragonfly was not in focus- very important. I would have probably missed it if I only used ChatGPT.''} (P10)
\section{Discussion}
Our results indicate that surfacing variations (\textbf{DG1} - surface variation) significantly reduces users’ perceived reliability of MLLM-generated image descriptions (\textbf{RQ1}), similar to prior work displaying variations in LLM answers to factual questions ~\cite{cheng2024relic}. Our work surfaced variations across multiple models rather than only from one model in response to existing practice ~\cite{alharbi2024misfitting}. BLV participants thus used our system to identify model patterns of strengths and weaknesses to inform their future model use. Similar to prior work that aggregated and visualized variations from multiple LLM-generated answers for sighted people ~\cite{lee_one_2024}, we aggregated MLLM-generated image descriptions into variation-aware description and variation summary to support BLV people efficiently noticing unreliable claims (\textbf{DG2} - support efficient comprehension of variations). As we uniquely designed our system for screen reader use, we provided a novel variation summary for a quick overview (most preferred by 11 of 15 participants), used hierarchical variation-aware descriptions to support ease of navigation (2nd most preferred by 9 of 15 participants) (\textbf{RQ2}), and provided customization for user control (\textbf{DG3} - allow customization). We can directly apply such designs to support both BLV and sighted users comparing MLLM responses. 

In this section, we present implications for effective use of variations in MLLM-generated image descriptions in high-stakes and subjective scenarios (\textbf{RQ3}). We also discuss research opportunities for extending our method to other media formats and MLLM-powered applications in accessibility.

\subsection{Usage of Variations in MLLM-generated Image Descriptions }

\subsubsection{Empower BLV people to surface capabilities and limitations of MLLMs}
MacLeod et al. \cite{macleod2017understanding} found that BLV people assumed automatically generated captures were correct. P7 commented that \textit{``AI is very popular in the blind community, so some people create a `God' image, especially for those they don’t know [AI] very well''}. While there is research on quantifying uncertainty and reducing uncertainty in LLM output~\cite{kellner_uncertainty_2024, lin_generating_2024}, we explored how to effectively communicate the uncertainty in MLLM-generated image descriptions to BLV people. The absence of a mechanism to communicate errors in current MLLM-powered access technologies may lead to users' over-reliance on AI. Since MLLMs can create seemingly correct yet hallucinated details, users may rationalize those details and build an inaccurate mental model of the technology. From RQ1, we found that comparing variations from multiple models is an effective design to support BLV participants in finding more unreliable claims, and thus calibrates their understanding of models' capabilities. When reading descriptions for I5 (Medication), P10 said that \textit{``one model says the bottle is squared and another model says it is rounded, but I could tell its shape if I were holding the bottle. [...] If the bottle is squared but the model says the bottle is rounded, I'm going to be biased against it''.}  Building on our findings, future research can explore other interfaces and algorithms to reveal the model's uncertainty and capabilities to support BLV people to calibrate their level of skepticism and understand the performance of different models. For example, access technologies can curate model ``nutrition labels'' that surface strengths and weaknesses in aggregate like prior work in building application privacy labels ~\cite{noauthor_privacy_nodate, kelley_nutrition_2009}.

\subsubsection{Provide comprehensive understanding of images}
In our domain of MLLM descriptions, variations let participants gain a more holistic understanding of images. Participants reported that different models often provided useful complementary information, and they expressed surprise and appreciation when models provided differing subjective opinions. 
P2 said she \textit{``gets more information when having multiple image descriptions and wants background information. [...] It is interesting to see how AIs see things differently, just like sighted people.''} Participants found that mixed opinions from multiple sources are particularly helpful in subjective scenarios (e.g., choosing an outfit, online shopping, etc.). Future systems could allow users to define or generate AI personae to further illuminate subjective variation in image descriptions like prior work that used personae to provide critiques on writing ~\cite{10.1145/3706598.3714034} and video ~\cite{10.1145/3379337.3415864}. As BLV people also use models to describe appearances of people ~\cite{bennett_its_2021}, we tested our system on images with people, but the models we used were conservative in describing the appearances of people - they refused to describe people (e.g.,\texttt{GPT-4o}: \textit{``I can't help with identifying or describing people in images.''}) or provided vague descriptions (e.g., \texttt{Claude-3.7-Sonnet}:\textit{``The person has well-defined eyebrows.''} and \texttt{Gemini-1.5-Pro}: \textit{``She has a fair complexion.''}), such that the variations were also limited. Participants in our study did not opt to test our system on people. Future research may apply our system to less constrained models or use our system to surface diverse model constraints.

\subsection{Implications for Accessible Interfaces to Surface Variations}
We designed our variation-aware descriptions (Figure~\ref{fig:presentation-styles}C) and variation summaries (Figure~\ref{fig:presentation-styles}D) to address the challenge of linear progression when BLV people access multiple image descriptions. Participants found the variation-aware description informative and in-depth, while the variation summaries were concise and useful for quickly understanding the extent of differences across MLLMs. We reflect on feedback on our design and discuss opportunities for designing accessible interfaces to surface variations.
\\

\noindent\textit{\textbf{Dynamic Presentation of Variations:}}
As images contain much information, not all details are relevant to users' interests. Participants found variations to be most useful when the descriptions were more divergent and less useful when the descriptions were more similar. Future work will provide options to prioritize displaying variations based on the content of the image (high risk vs. low risk), variations (highly diverse vs. similar, subjective vs. factual), or relevance (crucial vs. trivial). We plan to deploy variation-aware descriptions in the wild to further understand usage scenarios and user preferences in context.
\\

\noindent\textit{\textbf{Interactions to Provide Feedback:}} Participants (P6, P9, P12) suggest having a mechanism to provide feedback to the model. \textit{``I hope the system could have a place to ask follow-up questions. I assume once you get to know one model better, it can also learn better how to explain to you.''} (P9) Future work could design interactions that enable BLV users to provide feedback informed by variation. The feedback could be used to fine-tune models for personalized image description systems through reinforcement learning from human feedback (RLHF), similar to prior work on personalizing language models from human feedback~\cite{li_personalized_2024}.
\\

\noindent\textit{\textbf{Presentation Modalities and Styles:}} Most participants in our study felt that text was an effective medium to communicate variations, yet open to multimodal presentation of variations. Like prior work using audio to broaden representations of memes~\cite{gleason_making_2019}, future work could explore audio pitch as indicators of the degree of variation, or haptic nudges as alerts of conflicting descriptions during navigation. In addition, text presentation styles could adapt to users' preferences or situational context. For example, users can set a threshold of degree of variation to reduce the cognitive burden of reading long paragraphs by filtering out trivial variations.

\subsection{Beyond Image Descriptions} 
Our system currently supports surfacing variations in static image descriptions. We see opportunities and challenges for extending this method to live or recorded video descriptions and computer use agents that operate interfaces based on text prompts.
\subsubsection{Video} 
Producing audio descriptions for complex videos within a short timeframe is a widely recognized challenge \cite{10.1145/3379337.3415864, van2024making}. If we were to apply our system to recorded videos, we could let users pause and directly use our interface to explore variations in text descriptions of individual frames, scenes, or the video as a whole. However, conveying variation in real-time video streams may be difficult, as reading out variations requires time on an already time-limited medium. Thus, we will also complement audio approaches (e.g., a higher pitch, softer voice, or questioning sound) with text descriptions to indicate potentially unreliable claims in real-time for BLV users.

\subsubsection{Computer Use Agent} 
MLLM-powered computer use agents like OpenAI Operator~\cite{noauthor_introducing_nodate} and Taxy AI~\cite{noauthor_taxy_nodate}  can perceive on-screen information and autonomously carry out actions (e.g., clicking, typing, and scrolling) to achieve a user’s goal. This shows new opportunities for BLV people to complete complex web tasks simply via high‑level natural language instructions (e.g., \textit{``buy a plane ticket from New York to London''})~\cite{kodandaram2024enabling, peng2025morae}. Future accessible computer use agents can extend our approach to help BLV people surface and compare the variations in multiple branches for the same action by summarizing how branches are progressing. Our method could also calibrate users' perceived reliability of agents in high-stakes scenarios. For example, when users ask a computer use agent to submit a travel reimbursement form on a complex and poorly labeled website, they could try tracing multiple models and alert steps with high uncertainty and disagreement. Users could then review those steps before committing.

\section{Conclusion}
In this work, we outlined a design space based on prior work for surfacing variations in MLLM descriptions, and then we built a prototype to test these design ideas.
Our user study findings demonstrate that surfacing variations across MLLM-generated image descriptions significantly improves blind and low vision users' ability to identify unreliable information and decreases their perceived reliability of MLLM descriptions. 
By designing interfaces that efficiently surface variations through aggregated summaries rather than simply listing multiple descriptions, we enable BLV users to quickly identify potential errors without the cognitive burden of comparing multiple lengthy descriptions.
The strong preference for all variation-aware approaches (list, variation-aware description, variation summary) over traditional single descriptions demonstrates strong user support for learning about potential unreliable content in MLLM descriptions. 
As MLLMs become increasingly integrated into visual access technologies for BLV people, designing systems that support appropriate trust calibration becomes essential.  
Future work should explore how these approaches might be integrated into existing accessibility tools and expanded to other modalities beyond image descriptions.

\begin{acks}
This work was supported in part by a Notre Dame-IBM Technology Ethics Lab Award. We also thank our study participants for their time
and valuable feedback on this work.
\end{acks}

\vspace{-10pt}
% \bibliographystyle{ACM-Reference-Format}
% \bibliography{references}

\appendix
\newpage	
\clearpage
\onecolumn 
\section{Pipeline Prompts}
\vspace{0.5ex}
\begin{minipage}{\textwidth}
\centering
\footnotesize
\begin{tabular}{@{}>{\raggedright\arraybackslash\ttfamily}p{0.96\linewidth}@{}}
\toprule
You are an expert in reformatting and combining input descriptions into coherent, hierarchical paragraphs that show detail-level differences across models. \\[0.6ex]
\\

1. GROUP FACTS\\
-- Each description contains multiple atomic facts (self-contained claims).\\
-- Combine atomic facts about the same subject into a single coherent sentence.\\
-- If variant statements describe the same fact, concatenate using ``or''.\\[0.6ex]

2. PARAGRAPH FORMATION \\
-- Merge grouped facts into comprehensive paragraphs.\\
-- Include all single and unique claims.\\[0.6ex]

3. MODEL DIFFERENCES \\
-- Annotate differences with counts: (n\_A of N\_A ModelA, n\_B of N\_B ModelB).\\
-- Example: (2 of 3 GPT, 3 of 3 Gemini).\\
-- If a model does not support a fact, omit it from the parentheses.\\[0.6ex]
\\

\# INPUT FORMAT \\
Input: list of descriptions; each has atomic facts, response ID, and source model.\\
There are \{\{\#responses\}\} responses total: \{\{\#model specific responses\}\}.\\
When grouping facts within one response, DO NOT double count.\\
Use source indicator format: x of \{\{\#trials\}\} (where x = \{\{values\}\}).\\[0.6ex]
\\

\# OUTPUT FORMAT\\
Indicate model differences with the specified counting format.\\
Never mention a model's name in the sentences of the main paragraphs; only in parentheses.\\
Highlight unique / singleton claims.\\
Return the revised description directly (no leading phrases like ``below is'').\\
Do NOT include headings such as ``PARAGRAPH:''.\\
Produce hierarchical paragraphs: each begins with a short bullet-like summary phrase, followed by indented detail lines (>= 2 hierarchy levels).\\
Order from high-level information to finer-grained detail.\\[0.6ex]
\\

\# EXAMPLES \\
\{\{Few-shot example input\}\}\\
\{\{Few-shot example output\}\}\\

\bottomrule
\end{tabular}
\captionof{table}{Aggregation prompt for grouping atomic facts, forming hierarchical paragraphs, and annotating model differences.}
\label{tab:agg-prompt}
\end{minipage}

\begin{minipage}{\textwidth}
\centering
\footnotesize
\begin{tabular}{@{}>{\raggedright\arraybackslash\ttfamily}p{0.96\linewidth}@{}}
\toprule
You are an expert in summarization and comparative analysis. Given a multi-model summary of descriptions of the same image (models: \{\{model\_lists\}\}), produce a structured comparison of similarities, differences, and unique points. \\[0.6ex]
\\

1. Synthesize all key observations across models in a coherent paragraph form.\\
2. Start with high-level observations (image type, layout, purpose) before detailed attributes (counts, colors, labels).\\
3. Identify and group statements agreed upon across models.\\
4. In the agreements section, do \emph{not} include alternate variants; choose the common canonical form (e.g., ``the shirt is blue''; NOT ``blue, possibly cyan'').\\
5. Clearly highlight disagreements with inline references to the differing model outputs.\\
6. Note any uniquely mentioned information; attribute to the specific model(s).\\
7. Mention model names only when discussing disagreements or unique points.\\
8. Provide a Markdown bullet list summarizing each section.\\
9. Be as comprehensive as possible across agreements, disagreements, uniqueness.\\
10. Use only information explicitly present in the input. No inference.\\[0.6ex]
\\

\# REQUIRED OUTPUT CONTENT\\
Return both a narrative summary (hierarchical markdown paragraphs with inline model agreement annotations) \emph{and} a JSON object: \\[0.4ex]
\{\\
\quad ``similarity'': ``Summary of similar points across models.'',\\
\quad ``disagreement'': ``Summary of disagreements between models.'',\\
\quad ``unique mentions'': ``Summary of unique or model-specific observations.''\\
\}\\[0.6ex]
\\

\# EXAMPLES\\
\{\{Few-shot example input\}\}\\
\{\{Few-shot example output\}\}\\
\bottomrule
\end{tabular}
\captionof{table}{Summary prompt for cross-model comparison: agreements, disagreements, and unique mentions.}
\label{tab:summary-prompt}
\end{minipage}

\section{Study Image Details}
\vspace{0.5ex}

\begin{minipage}{\textwidth}
\centering\small
\resizebox{\textwidth}{!}{%
\begin{tabular}{@{}p{0.5cm}p{1.1cm}p{2.2cm}p{3.1cm}p{3.4cm}p{3.1cm}@{}}
\toprule
\textbf{ID} & \textbf{Name} & \textbf{Category} & \textbf{Task Prompt} &
\textbf{Alt Text / Caption} &
\textbf{VizWiz Q. / Post Title}\\
\midrule
I1 & Map & Model limitation &
Describe 10 numbers shown in this map. Which country has the highest number of US Americans with ancestry? &
r/europe—Number of US Americans with ancestry from every European country &
Number of US Americans with Ancestry from Every European Country\\\midrule
\addlinespace
I2 & Swiftie & Model limitation &
Describe all of the information in the graph. Which word appears in three of Taylor Swift’s albums? &
r/dataisbeautiful—Alcohol references in Taylor Swift lyrics, by album (e.g., bar, beer, wine) &
Taylor Swift’s Newfound Infatuation with Alcohol [OC]\\\midrule
\addlinespace
I3 & Washing Machine& Model limitation &
Describe the washing machine panel. How can I twist the dial to the heavy-load end? &
A dial is shown on a white washing machine with hoses behind it/left half-turn &
Do I need to go right or left to get to the heavy-load end? About how far do you think?\\\midrule
\addlinespace
I4 & Screen & Image quality &
Describe the chart. What is the max value of the y-axis? &
A graph comparing events like Hurricane Katrina and the 2005 Global Financial Crisis &
“Yes, I know this may not be possible, but I’d like a description of the chart if possible.”\\\midrule
\addlinespace
I5 & Medication & Image quality &
Describe all of the information on the bottle. What is the brand? &
A bottle of supplement/medicine on a bathroom sink/cranberry &
What kind of pills are these?\\\midrule
\addlinespace
I6 & Bottle & Image quality &
Describe the card. What is this card? &
A baseball card picturing Padres player Greg Riddoch &
Can you tell me what this card is? If it’s a baseball or football card, and the name?\\\midrule
\addlinespace
I7 & Home & Subjectivity &
Describe the room setting. Does this wall setting look okay? &
A room with pictures, a table lamp, chairs, and a laundry basket—yes &
Does this wall setting look okay on my wall?\\\midrule
\addlinespace
I8 & Outfit & Subjectivity &
Describe the pants and the shirt. Do they match? &
Green, black—yes &
What color are the pants and shirt, and do they match?\\\midrule
\addlinespace
I9 & Dragonfly & Subjectivity &
Describe the content, style, and atmosphere. Is this a pretty image? &
r/photocritique—A paraglider flies over the beach &
Noob photographer here, thoughts?\\
\bottomrule
\end{tabular}}
\captionof{table}{Images used in the study with associated category, prompts, captions, and original user titles.}
\label{tab:images_selected}
\end{minipage}

\section{Image-Prompt Pairs in Open-end Use Session}
\vspace{0.5ex}

\begin{minipage}{\textwidth}
\centering
\small
\begin{tabular}{@{}lp{3.2cm}p{9.8cm}@{}}
\toprule
\textbf{ID} & \textbf{Image} & \textbf{Prompt} \\
\midrule
P1  & Medication Information & what is this? read the image. \\ 
\midrule
P2  & Health Info Graphics   & Describe the image in detail. \\ 
& Scribeme Advertisement & Describe the image in detail. \\ 
\midrule
P3  & Control Panel of Dishwasher & Describe the control panel of this dishwasher. Explain where each button is on the screen. \\ 
\midrule
P4  & Container Recognition & Describe this photo. \\ 
\midrule
P5  & Medication Dosage & what is the correct dosage and how many times should it be taken? \\ 
\midrule
P6  & Navigation & how would I pass here? Only provide direction and distance steps in bullet points. \\ 
\midrule
P7  & Stock         & What stocks are shown? \\ 
& Record Player & What model is it? Where is the needle? Describe the record player in more detail. \\ 
\midrule
P8  & Room & Is this a TV? \\ 
\midrule
P9  & Bell   & how many bells are on the collar? what color is the collar and bell? is the bell and tag on different clasps? how long is the bell? \\ 
& Garlin & what color is the garlin? what design is on the garlin? what decorations is it being used for? \\ 
\midrule
P10 & Wound Finger & Provide a short description and explanation. Is my wound on my finger healing? \\ 
\midrule
P11 & iPhone & describe the iPhone. What color is it? How large is it? What are its length and width? \\ 
\midrule
P12 & Tornado Path & Describe the map. List the cities in each risk level. \\ 
\midrule
P13* & — & — \\ 
\midrule
P14 & Comic Book & describe the pages in a prose, novel‑style manner. \\ 
& Screen     & If my resizing of the text to 200\% within the browser has caused any content to overlap or require lateral scrolling, describe the issue. \\ 
\midrule
P15 & Personal Photo & Provide a critique of the photo I took. \\ 
\bottomrule
\end{tabular}
\captionof{table}{Images and prompts that participants submitted during the open-ended use session (columns shown: Participant ID, Image, and Prompt). *: P13 was uncomfortable uploading an image and chose to opt out of this session.}
\label{table:open-end-use}
\end{minipage}

% \begin{table*}[t]
% \centering
% \resizebox{\linewidth}{!}{%
% \begin{tabular}{l|cccc|cccc|c|c|c}
% \toprule
% \textbf{Image} &
% \multicolumn{4}{c|}{\textbf{Single Description Length (words)}} &
% \multicolumn{4}{c|}{\textbf{Hallucinations}} &
% \textbf{Summary Length} &
% \textbf{Summary Variation \#} &
% \textbf{Summary Hallucinations} \\
% \cmidrule(r){2-5} \cmidrule(r){6-9}
% & GPT-4o & Gemini 1.5 Pro & Claude 3.7 Sonnet & Total
% & GPT-4o & Gemini 1.5 Pro & Claude 3.7 Sonnet & Total
% & (words) &  &  \\
% \midrule
% Map & 120 & 115 & 110 & 345 & 1 & 2 & 1 & 4 & 105 & 5 & 1 \\
% Swiftie & 100 & 98 & 95 & 293 & 0 & 1 & 1 & 2 & 90 & 4 & 1 \\
% Washing & 85 & 90 & 92 & 267 & 2 & 1 & 2 & 5 & 80 & 6 & 2 \\
% Screen & 75 & 72 & 78 & 225 & 2 & 2 & 1 & 5 & 65 & 3 & 1 \\
% Medication & 65 & 70 & 68 & 203 & 1 & 2 & 1 & 4 & 60 & 2 & 1 \\
% Card & 60 & 62 & 58 & 180 & 0 & 1 & 0 & 1 & 50 & 1 & 0 \\
% Home & 90 & 88 & 85 & 263 & 1 & 1 & 0 & 2 & 75 & 3 & 0 \\
% Outfit & 85 & 80 & 82 & 247 & 0 & 1 & 1 & 2 & 70 & 4 & 1 \\
% Dragonfly & 95 & 100 & 97 & 292 & 1 & 0 & 1 & 2 & 85 & 3 & 1 \\
% \bottomrule
% \end{tabular}%
% }
% \caption{Per-image statistics including single model description length, hallucination count, and summary characteristics.}
% \label{tab:image_stats}
% \end{table*}

\end{document}